\documentclass[conference]{IEEEtran}

\usepackage[letterpaper, left=1in, right=1in, bottom=1in, top=0.75in]{geometry}

\usepackage{amsmath}
\usepackage{amssymb}
\usepackage{amsthm}
\usepackage{enumerate}
\usepackage{pgfplots}
\usepackage{tikz}
\usepackage{graphicx}
\usepackage{subfigure}
\usepackage{float}
\usepackage{adjustbox}
\usetikzlibrary{shapes,snakes}
\usetikzlibrary{calc}
\usetikzlibrary{patterns}
\usetikzlibrary{decorations.pathmorphing} 

\usepackage{array}
\usepackage{cite}
\usepackage{cleveref}

\usepackage{algorithm,algpseudocode}
\usepackage{algorithmicx}

\usepackage[utf8]{inputenc}

\newcolumntype{M}[1]{>{\centering\arraybackslash}m{#1}}
\newcolumntype{N}{@{}m{0pt}@{}}

\newtheorem{remark}{Remark}
\newtheorem{lemma}{Lemma}

\newcommand{\basestation}[3]{
\coordinate (a) at (#1,#2);
\draw[line width=(#3)*1.5pt,scale=#3] ($(a)+(0, -1.08)$) -- ($(a)+(0, 1)$);
\draw[line width=(#3)*1.5pt,scale=#3] ($(a)+(0,1)$) .. controls ($(a)+(-0.25,-0.5)$) .. ($(a)+(-0.5,-0.9)$);
\draw[line width=(#3)*1.5pt,scale=#3] ($(a)+(0,1)$) .. controls ($(a)+(0.25,-0.5)$) .. ($(a)+(0.5,-0.9)$);
\draw[line width=(#3)*1.5pt,scale=#3] ($(a)+(0, -0.9)$) -- ($(a)+(-0.35, -0.7)$);
\draw[line width=(#3)*1.5pt,scale=#3] ($(a)+(0, -0.9)$) -- ($(a)+(0.35, -0.7)$);
\draw[line width=(#3)*0.75pt,scale=#3] ($(a)+(-0.35, -0.65)$) -- ($(a)+(0, -0.5)$);
\draw[line width=(#3)*0.75pt,scale=#3] ($(a)+(0.35, -0.65)$) -- ($(a)+(0, -0.5)$);
\draw[line width=(#3)*1pt,scale=#3] ($(a)+(0, -0.6)$) -- ($(a)+(-0.3, -0.45)$);
\draw[line width=(#3)*1pt,scale=#3] ($(a)+(0, -0.6)$) -- ($(a)+(0.3, -0.45)$);
\draw[line width=(#3)*0.5pt,scale=#3] ($(a)+(-0.3, -0.45)$) -- ($(a)+(0, -0.32)$);
\draw[line width=(#3)*0.5pt,scale=#3] ($(a)+(0.3, -0.45)$) -- ($(a)+(0, -0.32)$);
\draw[line width=(#3)*0.75pt,scale=#3] ($(a)+(0, -0.3)$) -- ($(a)+(-0.22, -0.17)$);
\draw[line width=(#3)*0.75pt,scale=#3] ($(a)+(0, -0.3)$) -- ($(a)+(0.22, -0.17)$);
\draw[line width=(#3)*0.5pt,scale=#3] ($(a)+(-0.22, -0.17)$) -- ($(a)+(0, -0.07)$);
\draw[line width=(#3)*0.5pt,scale=#3] ($(a)+(0.22, -0.17)$) -- ($(a)+(0, -0.07)$);;
\draw[line width=(#3)*0.75pt,scale=#3] (a) -- ($(a)+(-0.18, 0.11)$);
\draw[line width=(#3)*0.75pt,scale=#3] (a) -- ($(a)+(0.18, 0.11)$);
\draw[line width=(#3)*0.5pt,scale=#3] ($(a)+(-0.18, 0.11)$) -- ($(a)+(0,0.2)$);
\draw[line width=(#3)*0.5pt,scale=#3] ($(a)+(0.18, 0.11)$) -- ($(a)+(0,0.2)$);   
\draw[line width=(#3)*0.5pt,scale=#3] ($(a)+(0, 0.3)$) -- ($(a)+(-0.1, 0.37)$);
\draw[line width=(#3)*0.5pt,scale=#3] ($(a)+(0, 0.3)$) -- ($(a)+(0.1, 0.37)$);
\draw[line width=(#3)*0.25pt,scale=#3] ($(a)+(-0.1, 0.37)$) -- ($(a)+(0, 0.43)$);
\draw[line width=(#3)*0.25pt,scale=#3] ($(a)+(0.1, 0.37)$) -- ($(a)+(0, 0.43)$);
\draw[line width=(#3)*0.75pt,scale=#3] ($(a)+(0, 1.2)$) -- ($(a)+(0,1)$);
\draw[fill=white,scale=#3] ($(a)+(0, 1.2)$) circle (0.05cm);
}
\newcommand{\basestationRed}[3]{
\coordinate (a) at (#1,#2);
\draw[line width=(#3)*1.5pt,scale=#3,color=red] ($(a)+(0, -1.08)$) -- ($(a)+(0, 1)$);
\draw[line width=(#3)*1.5pt,scale=#3,color=red] ($(a)+(0,1)$) .. controls ($(a)+(-0.25,-0.5)$) .. ($(a)+(-0.5,-0.9)$);
\draw[line width=(#3)*1.5pt,scale=#3,color=red] ($(a)+(0,1)$) .. controls ($(a)+(0.25,-0.5)$) .. ($(a)+(0.5,-0.9)$);
\draw[line width=(#3)*1.5pt,scale=#3,color=red] ($(a)+(0, -0.9)$) -- ($(a)+(-0.35, -0.7)$);
\draw[line width=(#3)*1.5pt,scale=#3,color=red] ($(a)+(0, -0.9)$) -- ($(a)+(0.35, -0.7)$);
\draw[line width=(#3)*0.75pt,scale=#3,color=red] ($(a)+(-0.35, -0.65)$) -- ($(a)+(0, -0.5)$);
\draw[line width=(#3)*0.75pt,scale=#3,color=red] ($(a)+(0.35, -0.65)$) -- ($(a)+(0, -0.5)$);
\draw[line width=(#3)*1pt,scale=#3,color=red] ($(a)+(0, -0.6)$) -- ($(a)+(-0.3, -0.45)$);
\draw[line width=(#3)*1pt,scale=#3,color=red] ($(a)+(0, -0.6)$) -- ($(a)+(0.3, -0.45)$);
\draw[line width=(#3)*0.5pt,scale=#3,color=red] ($(a)+(-0.3, -0.45)$) -- ($(a)+(0, -0.32)$);
\draw[line width=(#3)*0.5pt,scale=#3,color=red] ($(a)+(0.3, -0.45)$) -- ($(a)+(0, -0.32)$);
\draw[line width=(#3)*0.75pt,scale=#3,color=red] ($(a)+(0, -0.3)$) -- ($(a)+(-0.22, -0.17)$);
\draw[line width=(#3)*0.75pt,scale=#3,color=red] ($(a)+(0, -0.3)$) -- ($(a)+(0.22, -0.17)$);
\draw[line width=(#3)*0.5pt,scale=#3,color=red] ($(a)+(-0.22, -0.17)$) -- ($(a)+(0, -0.07)$);
\draw[line width=(#3)*0.5pt,scale=#3,color=red] ($(a)+(0.22, -0.17)$) -- ($(a)+(0, -0.07)$);;
\draw[line width=(#3)*0.75pt,scale=#3,color=red] (a) -- ($(a)+(-0.18, 0.11)$);
\draw[line width=(#3)*0.75pt,scale=#3,color=red] (a) -- ($(a)+(0.18, 0.11)$);
\draw[line width=(#3)*0.5pt,scale=#3,color=red] ($(a)+(-0.18, 0.11)$) -- ($(a)+(0,0.2)$);
\draw[line width=(#3)*0.5pt,scale=#3,color=red] ($(a)+(0.18, 0.11)$) -- ($(a)+(0,0.2)$);   
\draw[line width=(#3)*0.5pt,scale=#3,color=red] ($(a)+(0, 0.3)$) -- ($(a)+(-0.1, 0.37)$);
\draw[line width=(#3)*0.5pt,scale=#3,color=red] ($(a)+(0, 0.3)$) -- ($(a)+(0.1, 0.37)$);
\draw[line width=(#3)*0.25pt,scale=#3,color=red] ($(a)+(-0.1, 0.37)$) -- ($(a)+(0, 0.43)$);
\draw[line width=(#3)*0.25pt,scale=#3,color=red] ($(a)+(0.1, 0.37)$) -- ($(a)+(0, 0.43)$);
\draw[line width=(#3)*0.75pt,scale=#3,color=red] ($(a)+(0, 1.2)$) -- ($(a)+(0,1)$);
\draw[fill=white,scale=#3,color=red] ($(a)+(0, 1.2)$) circle (0.05cm);
}
\newcommand{\iPhone}[3]{
\coordinate (a) at (#1,#2);
\draw [line width=0.25pt,rounded corners=(#3)*1mm,fill=white,scale=(#3)] (a)--($(a)+(0.67,0)$)--($(a)+(0.67,1.381)$)--($(a)+(0,1.381)$)--cycle;
\draw [color=gray,line width=0.25pt,rounded corners=(#3)*0.8mm,fill=white,scale=(#3)] ($(a)+(0.015,0.015)$)--($(a)+(0.655,0.015)$)--($(a)+(0.655,1.366)$)--($(a)+(0.015,1.366)$)--cycle;
\draw [line width=0.25pt,rounded corners=(#3)*0.04mm,scale=(#3)] ($(a)+(0.2875,1.266)$)--($(a)+(0.3825,1.266)$)--($(a)+(0.3825,1.281)$)--($(a)+(0.2875,1.281)$)--cycle;
\draw[line width=0.25pt,scale=#3] ($(a)+(0.335,0.09)$) circle (0.055cm);
\draw[line width=0.25pt,scale=#3] ($(a)+(0.335,0.09)$) circle (0.044cm);
\draw[line width=0.25pt,scale=#3] ($(a)+(0.2275,1.2735)$) circle (0.015cm);
\draw[line width=0.25pt,scale=#3] ($(a)+(0.335,1.32)$) circle (0.01cm);
\draw [fill={rgb:black,1;white,4},line width=0.25pt,scale=(#3)] ($(a)+(0.042475,0.170195)$)--($(a)+(0.042475,0.170195)+(0.58505,0.0)$)--($(a)+(0.042475,0.170195)+(0.58505,1.04061)$)--($(a)+(0.042475,0.170195)+(0.0,1.04061)$)--cycle;
}
\newcommand{\iPhoneRed}[3]{
\coordinate (a) at (#1,#2);
\draw [line width=0.25pt,rounded corners=(#3)*1mm,fill=white,scale=(#3)] (a)--($(a)+(0.67,0)$)--($(a)+(0.67,1.381)$)--($(a)+(0,1.381)$)--cycle;
\draw [color=red,line width=0.25pt,rounded corners=(#3)*0.8mm,fill=white,scale=(#3)] ($(a)+(0.015,0.015)$)--($(a)+(0.655,0.015)$)--($(a)+(0.655,1.366)$)--($(a)+(0.015,1.366)$)--cycle;
\draw [line width=0.25pt,rounded corners=(#3)*0.04mm,scale=(#3)] ($(a)+(0.2875,1.266)$)--($(a)+(0.3825,1.266)$)--($(a)+(0.3825,1.281)$)--($(a)+(0.2875,1.281)$)--cycle;
\draw[line width=0.25pt,scale=#3] ($(a)+(0.335,0.09)$) circle (0.055cm);
\draw[line width=0.25pt,scale=#3] ($(a)+(0.335,0.09)$) circle (0.044cm);
\draw[line width=0.25pt,scale=#3] ($(a)+(0.2275,1.2735)$) circle (0.015cm);
\draw[line width=0.25pt,scale=#3] ($(a)+(0.335,1.32)$) circle (0.01cm);
\draw [fill={rgb:black,1;red,4},line width=0.25pt,scale=(#3)] ($(a)+(0.042475,0.170195)$)--($(a)+(0.042475,0.170195)+(0.58505,0.0)$)--($(a)+(0.042475,0.170195)+(0.58505,1.04061)$)--($(a)+(0.042475,0.170195)+(0.0,1.04061)$)--cycle;
}

\newcommand{\basestationGreen}[3]{
\coordinate (a) at (#1,#2);
\draw[line width=(#3)*1.5pt,scale=#3,color=green!50!black] ($(a)+(0, -1.08)$) -- ($(a)+(0, 1)$);
\draw[line width=(#3)*1.5pt,scale=#3,color=green!50!black] ($(a)+(0,1)$) .. controls ($(a)+(-0.25,-0.5)$) .. ($(a)+(-0.5,-0.9)$);
\draw[line width=(#3)*1.5pt,scale=#3,color=green!50!black] ($(a)+(0,1)$) .. controls ($(a)+(0.25,-0.5)$) .. ($(a)+(0.5,-0.9)$);
\draw[line width=(#3)*1.5pt,scale=#3,color=green!50!black] ($(a)+(0, -0.9)$) -- ($(a)+(-0.35, -0.7)$);
\draw[line width=(#3)*1.5pt,scale=#3,color=green!50!black] ($(a)+(0, -0.9)$) -- ($(a)+(0.35, -0.7)$);
\draw[line width=(#3)*0.75pt,scale=#3,color=green!50!black] ($(a)+(-0.35, -0.65)$) -- ($(a)+(0, -0.5)$);
\draw[line width=(#3)*0.75pt,scale=#3,color=green!50!black] ($(a)+(0.35, -0.65)$) -- ($(a)+(0, -0.5)$);
\draw[line width=(#3)*1pt,scale=#3,color=green!50!black] ($(a)+(0, -0.6)$) -- ($(a)+(-0.3, -0.45)$);
\draw[line width=(#3)*1pt,scale=#3,color=green!50!black] ($(a)+(0, -0.6)$) -- ($(a)+(0.3, -0.45)$);
\draw[line width=(#3)*0.5pt,scale=#3,color=green!50!black] ($(a)+(-0.3, -0.45)$) -- ($(a)+(0, -0.32)$);
\draw[line width=(#3)*0.5pt,scale=#3,color=green!50!black] ($(a)+(0.3, -0.45)$) -- ($(a)+(0, -0.32)$);
\draw[line width=(#3)*0.75pt,scale=#3,color=green!50!black] ($(a)+(0, -0.3)$) -- ($(a)+(-0.22, -0.17)$);
\draw[line width=(#3)*0.75pt,scale=#3,color=green!50!black] ($(a)+(0, -0.3)$) -- ($(a)+(0.22, -0.17)$);
\draw[line width=(#3)*0.5pt,scale=#3,color=green!50!black] ($(a)+(-0.22, -0.17)$) -- ($(a)+(0, -0.07)$);
\draw[line width=(#3)*0.5pt,scale=#3,color=green!50!black] ($(a)+(0.22, -0.17)$) -- ($(a)+(0, -0.07)$);;
\draw[line width=(#3)*0.75pt,scale=#3,color=green!50!black] (a) -- ($(a)+(-0.18, 0.11)$);
\draw[line width=(#3)*0.75pt,scale=#3,color=green!50!black] (a) -- ($(a)+(0.18, 0.11)$);
\draw[line width=(#3)*0.5pt,scale=#3,color=green!50!black] ($(a)+(-0.18, 0.11)$) -- ($(a)+(0,0.2)$);
\draw[line width=(#3)*0.5pt,scale=#3,color=green!50!black] ($(a)+(0.18, 0.11)$) -- ($(a)+(0,0.2)$);   
\draw[line width=(#3)*0.5pt,scale=#3,color=green!50!black] ($(a)+(0, 0.3)$) -- ($(a)+(-0.1, 0.37)$);
\draw[line width=(#3)*0.5pt,scale=#3,color=green!50!black] ($(a)+(0, 0.3)$) -- ($(a)+(0.1, 0.37)$);
\draw[line width=(#3)*0.25pt,scale=#3,color=green!50!black] ($(a)+(-0.1, 0.37)$) -- ($(a)+(0, 0.43)$);
\draw[line width=(#3)*0.25pt,scale=#3,color=green!50!black] ($(a)+(0.1, 0.37)$) -- ($(a)+(0, 0.43)$);
\draw[line width=(#3)*0.75pt,scale=#3,color=green!50!black] ($(a)+(0, 1.2)$) -- ($(a)+(0,1)$);
\draw[fill=white,scale=#3,color=green!50!black] ($(a)+(0, 1.2)$) circle (0.05cm);
}

\newcommand{\iPhoneGreen}[3]{
\coordinate (a) at (#1,#2);
\draw [line width=0.25pt,rounded corners=(#3)*1mm,fill=white,scale=(#3)] (a)--($(a)+(0.67,0)$)--($(a)+(0.67,1.381)$)--($(a)+(0,1.381)$)--cycle;
\draw [color=green!50!black,line width=0.25pt,rounded corners=(#3)*0.8mm,fill=white,scale=(#3)] ($(a)+(0.015,0.015)$)--($(a)+(0.655,0.015)$)--($(a)+(0.655,1.366)$)--($(a)+(0.015,1.366)$)--cycle;
\draw [line width=0.25pt,rounded corners=(#3)*0.04mm,scale=(#3)] ($(a)+(0.2875,1.266)$)--($(a)+(0.3825,1.266)$)--($(a)+(0.3825,1.281)$)--($(a)+(0.2875,1.281)$)--cycle;
\draw[line width=0.25pt,scale=#3] ($(a)+(0.335,0.09)$) circle (0.055cm);
\draw[line width=0.25pt,scale=#3] ($(a)+(0.335,0.09)$) circle (0.044cm);
\draw[line width=0.25pt,scale=#3] ($(a)+(0.2275,1.2735)$) circle (0.015cm);
\draw[line width=0.25pt,scale=#3] ($(a)+(0.335,1.32)$) circle (0.01cm);
\draw [fill={rgb:black,1;green!50!black,4},line width=0.25pt,scale=(#3)] ($(a)+(0.042475,0.170195)$)--($(a)+(0.042475,0.170195)+(0.58505,0.0)$)--($(a)+(0.042475,0.170195)+(0.58505,1.04061)$)--($(a)+(0.042475,0.170195)+(0.0,1.04061)$)--cycle;
}

\newcommand{\TxAntenna}[3]{
	\coordinate (a) at (#1,#2);
	\draw[line width=0.25pt,scale=(#3)] (a)--($(a)+(0.2,0)$)--($(a)+(0.2,0.7)$)--
	($(a)+(0.1,0.8)$)--($(a)+(0.3,0.8)$)--($(a)+(0.2,0.7)$);
}

\newcommand{\RxAntenna}[4]{
	\coordinate (a) at (#1,#2);
	\draw[line width=0.25pt,scale=(#3)] (a)--($(a)+(-0.2,0)$)--($(a)+(-0.2,0.7)$)--
	($(a)+(-0.1,0.8)$)--($(a)+(-0.3,0.8)$)--($(a)+(-0.2,0.7)$);
}

\newcommand{\anas}{\textcolor{black}}
\begin{document}
%
\title{On The Efficiency of Widely Linear Precoding and Symbol Extension in Cellular Uplink}

\author{\IEEEauthorblockN{Ali Kariminezhad, Stefan Roth, Aydin Sezgin}
\IEEEauthorblockA{Digital Communication Systems\\Ruhr University Bochum\\
Email: \{ali.kariminezhad,stefan.roth-k21,aydin.sezgin\}@rub.de}
\and
\IEEEauthorblockN{Anas Chaaban}
\IEEEauthorblockA{School of Engineering\\ University of British Columbia\\
Email: anas.chaaban@ubc.ca}}


%


\maketitle

\begin{abstract}
We investigate Gaussian widely linear precoding known as improper Gaussian signaling for the cellular uplink with inter-cell interference, known as interference multiple access channel (IMAC). This transmission scheme provides extra degrees of freedom by treating the real and imaginary components of the complex Gaussian signal differently. Since current standards mainly utilize linear beamforming for waveform generation, we highlight the benefits of widely linear beamforming over multiple temporal dimensions (symbol extension in time) in the IMAC. This scheme achieves significantly higher information rates compared to conventional proper Gaussian signaling at the expense of extra complexity at the transmission phase. We study the sum-power minimization problem under rate constraints. This problem is a difference of concave functions (DC) program, hence, a non-convex problem. By numerical simulations, we observe the benefits of improper Gaussian signaling alongside symbol extension in power consumption for both single-antenna and multi-antenna base stations. Interestingly, we observe that at strong interference scenarios, the efficiency of improper Gaussian signaling outperforms conventional proper Gaussian signaling at low rate demands. Moreover, in such scenarios the sum-power required for achieving particular rate demands is significantly reduced.

\end{abstract}

\section{Introduction}
An increase in the number of users in future communication systems is inevitable~\cite{Cisco}. In the context of cellular communication, a plethora of greedy users will coexist in multiple cells, all of which are demanding reliable communication with high data rates. As the number of users increases, the probability of simultaneous transmission requests increases. Dividing the resources (time and bandwidth) among users for interference-free access can not sustain this load, since each user will only get a small portion of the overall network resources, not enough to achieve the desired performance. At this point, resource-sharing becomes necessary. 
\begin{figure}
\centering
\tikzset{every picture/.style={scale=.65}, every node/.style={scale=0.9}}%
\begin{tikzpicture}
\draw[fill=yellow!80!red,opacity=0.3] (0,0) ellipse (3cm and 2cm);
\basestation{0}{0}{0.8};
\iPhone{-2}{0}{0.4};
\iPhone{1.5}{-1}{0.4};
\iPhone{1.5}{1}{0.4};

\draw[fill=green!80!red,opacity=0.3] (2,-2) ellipse (3cm and 2cm);
\basestationRed{2}{-2}{0.8};
\iPhoneRed{0}{-1.7}{0.4};
\iPhoneRed{3}{-3.5}{0.4};
\iPhoneRed{4}{-2.5}{0.4};

\draw[fill=blue!80!red,opacity=0.3] (-2,-2) ellipse (3cm and 2cm);
\basestationGreen{-2}{-2}{0.8};
\iPhoneGreen{-1}{-1.7}{0.4};
\iPhoneGreen{-2}{-3.5}{0.4};
\iPhoneGreen{-4}{-2}{0.4};
\end{tikzpicture}
\caption{\small Interference multiple access channel (IMAC). Multiple users are accessing the base stations exploited in multiple cells, while causing inter-cell interference. Serving base stations are distinguished by different colors.}
\label{fig:SystemModel}
\end{figure}
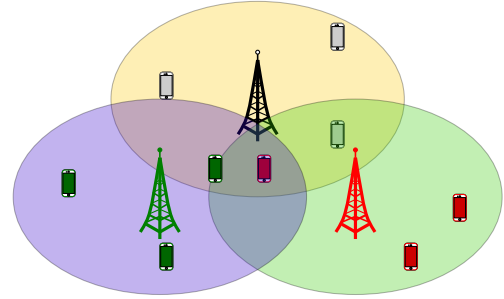

Resource-sharing in time and frequency increases interference. This requires smart interference management strategies at the cost of transceiver complexity. Generally, higher degrees of freedom in designing the transmit signal allow for better interference management capabilities. Here, we define degrees of freedom (DoF) as the number of independent interference-free streams, that are decoded with arbitrarily small error rate. DoF approximates the channel capacity at very high signal-to-noise ratio (SNR). In a time-variant $K$-user interference channel, $\frac{1}{2}$ (DoF) per-user is achievable using an interference management scheme known as interference alignment (IA)~\cite{Jafar2008}. This is significantly better than orthogonal resource allocation, e.g., TDMA, FDMA, where only $\frac{1}{K}$ DoF per-user can be obtained. This strong result inspired the authors of~\cite{Anas2011B} to study the DoF of the partial interference multiple access channel (PIMAC). The authors in~\cite{Jafar2009} investigated the DoF of the 3-user time-invariant interference channel (IC). They showed that by improper Gaussian signaling (IGS) over multiple temporal dimensions (an extended symbol in time) and interference alignment (IA) a sum-degrees of freedom (sum-DoF) of $\frac{6}{5}$ is achievable, which is again higher than sum-DoF achievable of 1 by orthogonal resource allocation procedures. Recall that these results describe the performance of the transmission schemes at very high SNR. Hence, it is of interest to investigate the performance of these schemes at low/moderate SNR.

In the low/moderate SNR regime, sacrificing signal dimensions for aligning the interference is not necessarily the optimal strategy. Hence, depending on the SINR, the signal space can be exploited more efficiently in order to optimize utility. Additionally, transmission power is an essential performance criterion in this regime, not only transmission rate. Due to the fact that IGS includes PGS as a special case (uncorrelated real and imaginary components with equal power), IGS always performs always better or at least as good as PGS at SINR from both rate and power perspectives. The authors of~\cite{Ho2012} show the benefits of IGS in 2-user IC in terms of achievable rates and the authors of~\cite{Zeng2013} investigate the achievable rate region of IGS in a $K$-user IC. The authors of~\cite{KariminezhadICC2017} highlight the power efficiency of IGS in MIMO full-duplex relaying for $K$-user interference networks. The rate-energy region of a two-tier network is investigated in~\cite{KariminezhadS16}. Moreover, the efficiency of IGS alongside symbol extension is studied from the energy efficiency perspective in~\cite{KariminezhadCS16}. The authors of~\cite{Fritschek2014},\cite{PangV16} study the generalized degrees of freedom (GDoF) of deterministic and Gaussian IMAC. Moreover, the GDoF region of the partial IMAC is investigated in~\cite{Gherekhloo2017} and the achievable rate region of the partial IMAC is studied in~\cite{Kariminezhad2017C}.

In this paper, we investigate an uplink channel in multiple adjacent cells sharing the same resources. In such a channel, the desired signals within a cell suffer from the inter-cell interference from the neighboring cells. This channel is called an interference multiple-access channel (IMAC) throughout the paper, Fig.~\ref{fig:SystemModel}. 

Exploiting IGS over an extended symbol, we investigate the power consumption of the IMAC. We formulate a power minimization problem under rate constraints. The obtained optimization problem turns out to be a difference of convex (DC) program. \anas{We design an algorithm which is based on successive convex approximation of the non-convex constraint set, where the approximation gap is reduced iteratively. We evaluate the solution numerically, and interestingly, we observe that by IGS and symbol extension, the required power for achieving target rates is significantly reduced.} For instance, in strong interference scenarios, almost quarter the sum-power of  PGS is required by IGS to achieve 0.6 bit per channel use (bit/cu) over an extended symbol of length 2.
\subsection{Notation} Throughout the paper, we represent vectors using boldface lower-case letters and matrices using boldface upper-case letters. ${\rm{Tr}}(\bf{A})$, $|{\bf{A}}|$, ${\bf{A}}^{H}$, ${\bf{A}}^{T}$, ${\bf{A}}^{-1}$ represent the trace, determinant, hermitian, transpose and inverse of matrix $\mathbf A$, respectively. ${\bf I}_N$ denotes the identity matrix of size $N$. The notation $\otimes$ represents Kronecker product of two matrices. The cardinality of set $\mathcal{A}$ is represented by $|\mathcal{A}|$. Real and imaginary components of $x$ are denoted by $\Re({x})$ and $\Im({x})$, respectively.

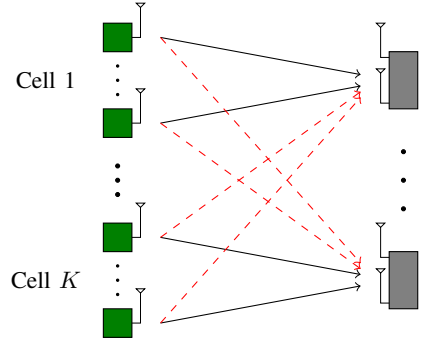
\begin{figure}
\centering
\tikzset{every picture/.style={scale=.76}, every node/.style={scale=0.9}}%
\begin{tikzpicture}[scale=0.5]
\node at (-2,-1){Cell 1};
\draw[fill=green!50!black] (0,0) rectangle (1,1);
\TxAntenna{1}{0.5}{1.5};
\draw[fill] (0.5,-0.5) circle (0.04cm);
\draw[fill] (0.5,-1) circle (0.04cm);
\draw[fill] (0.5,-1.5) circle (0.04cm);
\draw[fill=green!50!black] (0,-3) rectangle (1,-2);
\TxAntenna{1}{-2.5}{1.5};

\draw[fill] (0.5,-4) circle (0.07cm);
\draw[fill] (0.5,-4.5) circle (0.07cm);
\draw[fill] (0.5,-5) circle (0.07cm);

\node at (-2,-8){Cell $K$};
\draw[fill=green!50!black] (0,-7) rectangle (1,-6);
\TxAntenna{1}{-6.5}{1.5};
\draw[fill] (0.5,-7.5) circle (0.04cm);
\draw[fill] (0.5,-8) circle (0.04cm);
\draw[fill] (0.5,-8.5) circle (0.04cm);
\draw[fill=green!50!black] (0,-10) rectangle (1,-9);
\TxAntenna{1}{-9.5}{1.5};

\draw[fill=gray] (10,-2) rectangle (11,0);
\RxAntenna{10}{-1.8}{1.5};
\RxAntenna{10}{-0.2}{1.5};
\draw[fill] (10.5,-3.5) circle (0.07cm);
\draw[fill] (10.5,-4.5) circle (0.07cm);
\draw[fill] (10.5,-5.5) circle (0.07cm);
\draw[fill=gray] (10,-9) rectangle (11,-7);
\RxAntenna{10}{-8.8}{1.5};
\RxAntenna{10}{-7.2}{1.5};

\draw[->] (2,0.5)--(9,-0.8);
\draw[->] (2,-2.5)--(9,-1.2);

\draw[->] (2,-6.5)--(9,-7.8);
\draw[->] (2,-9.5)--(9,-8.2);

\draw[->,dashed,red] (2,0.5)--(9,-7.4);
\draw[->,dashed,red] (2,-2.5)--(9,-7.6);

\draw[->,dashed,red] (2,-6.5)--(9,-1.4);
\draw[->,dashed,red] (2,-9.5)--(9,-1.6);
\end{tikzpicture}
\caption{\small Interference multiple access channel (IMAC).}
\label{fig:SystemModelB}
\end{figure}

\section{System Model}
We consider a cellular network, where multiple single-antenna mobile stations (MS) are located in $K$ cells. Each cell is equipped with an access-point with $M$ antennas, as shown in~Fig.~\ref{fig:SystemModelB}. We denote the complex-valued transmit signal from the $j$th user in the $k$th cell by $x_{j_k}$. Then, the received signal at the $k$th access-point is given by 
\begin{align}
\mathbf{y}_k=\sum_{j=1}^{|\mathcal{I}_k|}\mathbf{h}_{kj_k}x_{j_k}+\sum_{\substack{l=1\\l\neq k}}^{K}\sum_{j=1}^{|\mathcal{I}_l|}\mathbf{h}_{kj_l}x_{j_l}+\mathbf{w}_k,
\end{align}
where the set of MSs at the $k$th cell is represented by $\mathcal{I}_k$, so that the cardinality of this set represents the number of users in that cell.  The channel from the $j$th MS located in the $l$th cell to the BS in the $k$th cell is depicted by $\mathbf{h}_{kj_l}\in\mathbb{C}^{M}$, which is globally known and is assumed to have sufficiently large coherence time. The receiver additive noise at the $k$th BS is represented by $\mathbf{w}_k\in\mathbb{C}^{M}$, which is assumed to follow proper Gaussian distribution with mean zero and covariance $\sigma^2_k\mathbf{I}_M$, i.e., $\mathbf{w}_k\sim\mathcal{CN}(\mathbf{0},\sigma^2_k\mathbf{I}_M)$. The transmit $x_{j_k}$ is from a Gaussian codebook with power $p_{j_k}$. We denote the $i$th component of a vector $\mathbf{x}$ by $x^{(i)}$.

Now, we represent the complex-valued equivalent SISO channel, by its real-valued model. This can be done by stacking the real and imaginary components of the transmit and received signal into vectors. Hence, we define
\begin{align}
\hat{\mathbf{y}}_{k}=&[\Re\left(y^{(1)}_{k}\right)\quad \Im\left(y^{(1)}_{k}\right),\cdots\nonumber\\
&\Re\left(y^{(M)}_{k}\right)\quad \Im\left(y^{(M)}_{k} \right)]^{T},\\
\hat{\mathbf{x}}_{i_k}=&[\Re\left(x_{i_k}\right)\quad \Im\left(x_{i_k}\right)]^{T},\\
\hat{\mathbf{w}}_{k}=&[\Re\left(w^{(1)}_{k}\right)\quad \Im\left(w^{(1)}_{k}\right),\cdots\nonumber\\
&\Re\left(w^{(M)}_{k}\right)\quad \Im\left(w^{(M)}_{k} \right)]^{T},
\end{align}
Then, the real-valued equivalent channel input-output relationship is formulated as
\begin{align}
\hat{\mathbf{y}}_{k}=\sum_{j=1}^{|\mathcal{I}_k|}\mathbf{G}_{kj_k}\hat{\mathbf{x}}_{j_k}+\sum_{\substack{l=1\\l\neq k}}^{K}\sum_{j=1}^{|\mathcal{I}_l|}\mathbf{G}_{kj_l}\hat{\mathbf {x}}_{j_l}+\hat{\mathbf{w}}_{k},
\end{align}
where the real-valued equivalent channel matrix is given by

{\small
\begin{align}
\mathbf{G}_{kj_l}=\begin{bmatrix}
\Re{(h^{(1)}_{kj_l})}&-\Im{(h^{(1)}_{kj_l})}\\
\Im{(h^{(1)}_{kj_l})}&\Re{(h^{(1)}_{kj_l})}\\ \vdots & \vdots \\
\Re{(h^{(M)}_{kj_l})}&-\Im{(h^{(M)}_{kj_l})}\\
\Im{(h^{(M)}_{kj_l})}&\Re{(h^{(M)}_{kj_l})}\\
\end{bmatrix}.
\end{align}}
Suppose that, the channel is time-invariant over $N$ time instants. Then, the received signal vector over these $N$ time instants is given by
\begin{align}
\bar{\mathbf{y}}_{k}=\sum_{j=1}^{|\mathcal{I}_k|}\bar{\mathbf{G}}_{kj_k}\bar{\mathbf{x}}_{i_k}+\sum_{\substack{l=1\\l\neq k}}^{K}\sum_{j=1}^{|\mathcal{I}_l|}\bar{\mathbf{G}}_{kj_l}\bar{\mathbf {x}}_{j_l}+\bar{\mathbf{w}}_{k},\label{MimoA}
\end{align}
where $\bar{\mathbf{G}}_{kj_l}=\mathbf{I}_N\otimes \mathbf{G}_{kj_l}$. Notice that, $\bar{\mathbf{y}}_{k}$ and $\bar{\mathbf{w}}_{k}$ stacks $N$ time samples of the received signal and receiver noise vectors into single vectors, respectively. Moreover, $\bar{\mathbf{x}}_{i_k}$ precodes the real-valued transmit signal vector over $N$ channel uses, i.e., symbol extension of length $N$. 
In the real-valued equivalent MIMO channel represented in~\eqref{MimoA}, the achievable rate for the message of the $i$th user in the $k$th cell, denoted by $R_{i_k}$, is bounded as shown in~\eqref{MimoRate} at the top of the next page~\cite{Telatar1999}. In~\eqref{MimoRate}, the transmit covariance matrix of the $i$th MS in the $k$th cell is denoted by $\mathbf{Q}_{i_k}$, i.e., $\mathbf{Q}_{i_k}=\mathbb{E}\{{\bar{\mathbf{x}}_{i_k}\bar{\mathbf{x}}^{H}_{i_k}}\}$ . This covariance matrix captures the joint design is signal-space and time. \anas{As can be noticed in~\eqref{MimoRate}, the base stations perform successive decoding (SD), while decoding the signals of the users.}
\begin{figure*}
{\small
\begin{align}
R_{i_k}\leq&\log_2\Big|\frac{\sigma^2_k}{2}\mathbf{I}_{2MN}+\sum_{j=	i}^{|\mathcal{I}_k|}\bar{\mathbf{G}}_{kj_k}\mathbf{Q}_{j_k}\bar{\mathbf{G}}^{H}_{kj_k}+\sum_{\substack{l=1\\l\neq k}}^{K}\sum_{j=1}^{|\mathcal{I}_l|}\bar{\mathbf{G}}_{kj_l}\mathbf{Q}_{j_l}\bar{\mathbf{G}}^{H}_{kj_l}\Big|\nonumber\\
&-\log_2\Big|\frac{\sigma^2_k}{2}\mathbf{I}_{2MN}+\sum_{j=i+1}^{|\mathcal{I}_k|}\bar{\mathbf{G}}_{kj_k}\mathbf{Q}_{j_k}\bar{\mathbf{G}}^{H}_{kj_k}+\sum_{\substack{l=1\\l\neq k}}^{K}\sum_{j=1}^{|\mathcal{I}_l|}\bar{\mathbf{G}}_{kj_l}\mathbf{Q}_{j_l}\bar{\mathbf{G}}^{H}_{kj_l}\Big|\label{MimoRate}
\end{align}}
\hrule
\end{figure*}

Our goal is to minimize the transmit power subject to target rates for the users. In the next section we formulate the sum-power minimization problem.

\section{Power minimization problem}
Consider that, the users in all cells have particular quality of service (QoS) demands. Then, it is of crucial importance to fulfill these demands by efficient transceiver design.
In this paper, the QoS demands are reflected by information rates. Hence, the sum-power minimization under rate demands is cast as
\begin{subequations}\label{OP:A}
\begin{align}
\min_{\mathbf{Q}_{i_k},\forall i\in\mathcal{I}_k, k\in\mathcal{K}}\quad& \sum_{k=1}^{K}\sum_{i=1}^{|\mathcal{I}_k|}\text{Tr}\left(\mathbf{Q}_{i_k}\right)\tag{\ref{OP:A}}\\
\text{subject to}\quad & \bar{R}_{i_k}\geq\psi_{i_k},\ \forall i,k\label{OP:A1}\\
& \text{Tr}\left(\mathbf{Q}_{i_k}\right)\leq P_{i_k},\ \forall i,k\label{OP:A2}\\
&\mathbf{Q}_{i_k}\succeq 0,\ \forall i,k \label{OP:A3}\\
&\mathbf{Q}_{i_k}\in\mathbb{S}^{2N\times 2N},\ \forall i,k, \label{OP:A4}
\end{align}
\end{subequations}
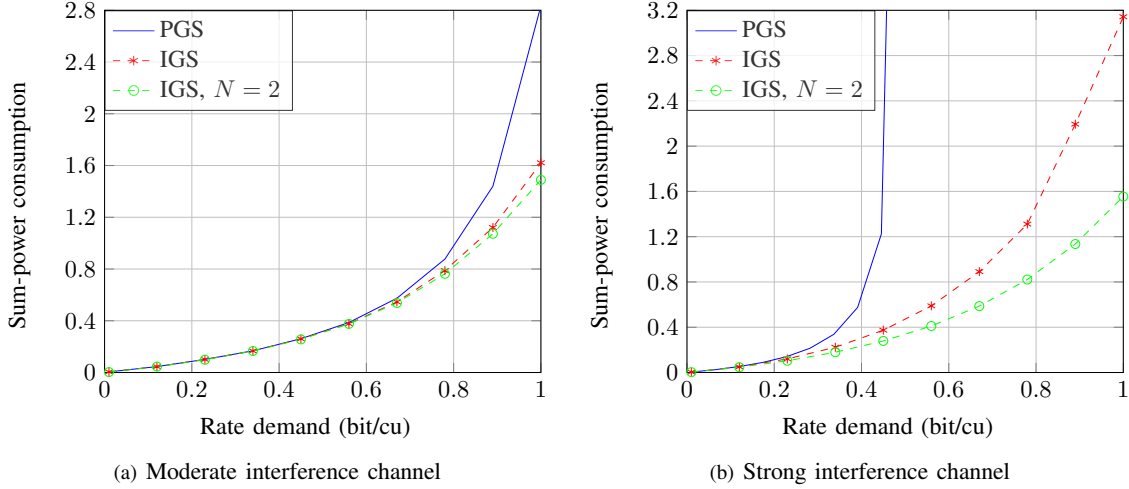
\begin{figure*}[h!]
\centering
\subfigure[\small Moderate interference channel]{
\tikzset{every picture/.style={scale=0.92}, every node/.style={scale=1}}%
\begin{tikzpicture}

\begin{axis}[%
xmin=0,
xmax=1,
xtick={0,0.2,0.4,0.6,0.8,1},
xlabel={Rate demand (bit/cu)},
xmajorgrids,
ymin=0,
ymax=2.8,
ytick={0,0.4,0.8,1.2,1.6,2,2.4,2.8},
ylabel={Sum-power consumption},
ymajorgrids,
legend style={at={(axis cs: 0,2.8)},anchor=north west,draw=black,fill=white,legend cell align=left,opacity=0.8}
]
\addplot [color=blue,solid]
  table[row sep=crcr]{0.01	0.00497744971669773\\
  0.12	0.0456290327003392\\
  0.23	0.101393358257957\\
  0.34	0.167448938703006\\
  0.45	0.261127456254858\\
  0.56	0.386485573059829\\
  0.67	0.574435280514849\\
  0.78	0.876914807576372\\
  0.89	1.44016977795369\\
  1	2.83461206880221\\
  };
\addlegendentry{PGS};

\addplot [color=red,dashed,mark=asterisk,mark options={solid}]
  table[row sep=crcr]{0.01	0.00491406494613247\\
  0.12	0.0455765297211817\\
  0.23	0.101018850638897\\
  0.34	0.166935517942434\\
  0.45	0.260243909602643\\
  0.56	0.3769044724385\\
  0.67	0.545821918356623\\
  0.78	0.788070655888636\\
  0.89	1.12114048713716\\
  1	1.62114938665051\\
  };
\addlegendentry{IGS};

\addplot [color=green,dashed,mark=o,mark options={solid}]
  table[row sep=crcr]{0.01	0.00340832367295722\\
  0.12	0.0454865964719718\\
  0.23	0.0986213701221813\\
  0.34	0.166603779489687\\
  0.45	0.255891205891875\\
  0.56	0.375645178969585\\
  0.67	0.537702331452102\\
  0.78	0.762915857465391\\
  0.89	1.07229757119441\\
  1	1.48924276660763\\
  };
\addlegendentry{IGS, $N=2$};

\end{axis}
\end{tikzpicture}%
\label{fig:SisoWeakInterference}
}
\subfigure[\small Strong interference channel]{
\tikzset{every picture/.style={scale=0.92}, every node/.style={scale=1}}%
\begin{tikzpicture}

\begin{axis}[%
xmin=0,
xmax=1,
xtick={0,0.2,0.4,0.6,0.8,1},
xlabel={Rate demand (bit/cu)},
xmajorgrids,
ymin=0,
ymax=3.2,
ytick={0,0.4,0.8,1.2,1.6,2,2.4,2.8,3.2},
ylabel={Sum-power consumption},
ymajorgrids,
legend style={at={(axis cs: 0,3.2)},anchor=north west,draw=black,fill=white,legend cell align=left,opacity=0.8}
]
\addplot [color=blue,solid]
  table[row sep=crcr]{0.01	0.00404194208664495\\
  0.0644444444444444	0.024863527627964\\
  0.118888888888889	0.0522969682987152\\
  0.173333333333333	0.0888095059968116\\
  0.227777777777778	0.139386412562958\\
  0.282222222222222	0.214991265614085\\
  0.336666666666667	0.337592222018465\\
  0.391111111111111	0.575094926819577\\
  0.445555555555556	1.2248446637732\\
  0.5	10.2465902600043\\
  };
\addlegendentry{PGS};

\addplot [color=red,dashed,mark=asterisk,mark options={solid},]
  table[row sep=crcr]{0.01	0.00343320967947817\\
  0.12	0.0503152036267284\\
  0.23	0.120125196175928\\
  0.34	0.223511397054854\\
  0.45	0.373893584142563\\
  0.56	0.588847644957806\\
  0.67	0.891141113735411\\
  0.78	1.31116732985599\\
  0.89	2.19416461159308\\
  1	3.14278689817075\\
  };
\addlegendentry{IGS};

\addplot [color=green,dashed,mark=o,mark options={solid}]
  table[row sep=crcr]{0.01	0.00342538054272735\\
  0.12	0.0468548527512783\\
  0.23	0.103939145399038\\
  0.34	0.179261123315517\\
  0.45	0.278845255485986\\
  0.56	0.411019326639699\\
  0.67	0.586844612326973\\
  0.78	0.821474231858168\\
  0.89	1.13510610164935\\
  1	1.55520010726407\\
  };
\addlegendentry{IGS, $N=2$};

\end{axis}
\end{tikzpicture}%
\label{fig:SisoStrongInterference}
}
\caption{\small Two cells, two users per cell, one antennas at the base station. Minimum power required to fulfill certain rate demands which is assumed to be equal for all users. Rate demands for all users are assumed to be equal. We assume that the channels remain constant over two symbols. The symbol extension length is represented by $N$.}
\label{fig:SISO}
\end{figure*} 
where the power budget at the $i$th MS in the $k$th cell is represented by $P_{i_k}$. Notice that the achievable rate bound in~\eqref{MimoRate} is denoted by $\bar{R}_{i_k}$. Moreover, the set of $2N\times 2N$ symmetric matrices is depicted by $\mathbb{S}^{2N\times 2N}$. Notice that, $\mathbf{Q}_{i_k},\ \forall i,\forall k$ are real-valued covariance matrices, hence, they should be symmetric positive semidefinite. $\psi_{i_k}$ represents the information rate demand of the $i$th MS in the $k$th cell. These demands might not be satisfied by the available resources, which renders the demands to be infeasible.
\begin{remark}
The QoS demands of the users might not be feasible by classical PGS. However, IGS over an extended symbol makes more efficient use of the transmit power budget which could satisfy the rate demands.
\end{remark}

The utility function in the optimization problem~\eqref{OP:A} is an affine function, however, it has a non-convex constraint~\eqref{OP:A1}, contrary to~\eqref{OP:A2}-\eqref{OP:A4} which are convex. This is due to the fact that, $\bar{R}_{i_k},\ \forall i,k,$ are the difference between concave functions as in~\eqref{MimoRate}. This makes the sum-power minimization problem a difference of concave functions (DC) program. Obtaining a good sub-optimal solution of a DC program in a polynomial time is a difficult task, in general. In this paper, we exploit an iterative algorithm to obtain an efficient sub-optimal solution.  Recall that, the two log-determinant functions in~\eqref{MimoRate} are concave in $\mathbf{Q}_{i_k}$. By linearizing the second term the whole expression becomes a concave function in $\mathbf{Q}_{i_k}$. Defining the received signal and the interference-plus-noise covariance matrices as
\begin{align}
\mathbf{A}_{i_k}=&\frac{\sigma^2_k}{2}\mathbf{I}_{2MN}+\sum_{j=i}^{|\mathcal{I}_k|}\bar{\mathbf{G}}_{kj_k}\mathbf{Q}_{j_k}\bar{\mathbf{G}}^{H}_{kj_k}\nonumber\\
+&\sum_{\substack{l=1\\l\neq k}}^{K}\sum_{j=1}^{|\mathcal{I}_l|}\bar{\mathbf{G}}_{kj_l}\mathbf{Q}_{j_l}\bar{\mathbf{G}}^{H}_{kj_l},\label{11A}\\
\mathbf{B}_{i_k}=&\mathbf{A}_{i_k}-\bar{\mathbf{G}}_{ki_k}\mathbf{Q}_{i_k}\bar{\mathbf{G}}^{H}_{ki_k}\label{11B}
\end{align}
respectively, and exploiting Fenchel's inequality and the concept of the conjugate function, we obtain the following upper-bound~\cite{Boyd2004}
\begin{align}
&\log_2|\mathbf{B}_{i_k}|\leq\log_2|{\boldsymbol\Gamma}_{i_k}|+{\rm Tr}({\boldsymbol\Gamma}_{i_k}^{-1}\mathbf{B}_{i_k})-MN,\label{FenchelA}
\end{align}
where the auxiliary matrix variables $\boldsymbol{\Gamma}_{i_k},\forall i,k$. The upper-bound gap in~\eqref{FenchelA} closes at optimal $\boldsymbol{\Gamma}_{i_k},\forall i,k$, which is $\boldsymbol{\Gamma}^{\star}_{i_k}=\mathbf{B}_{i_k}$.
Exploiting this upper-bound, the achievable rates bound i.e., $\bar{R}_{i_k}$, is lower-bounded by
\begin{align}
\bar{R}_{i_k}&=\log_2|\mathbf{A}_{i_k}|-\log_2|\mathbf{B}_{i_k}|\nonumber\\
&\geq\log_2|\mathbf{A}_{i_k}|-\log_2|{\boldsymbol\Gamma}_{i_k}|-{\rm Tr}({\boldsymbol\Gamma}_{i_k}^{-1}\mathbf{B}_{i_k})+MN\nonumber\\&:=
\tilde{R}_{i_k},\label{LBa}
\end{align}
where given any $\boldsymbol{\Gamma}_{i_k}$, the rates upper-bound $\bar{R}_{i_k}$ is lower-bounded by a concave expression in $\mathbf{Q}_{i_k}$. Now, the optimization problem~\eqref{OP:A} is reformulated as
\begin{subequations}\label{OP:B}
\begin{align}
\min_{\boldsymbol{\Gamma}_{i_k},\mathbf{Q}_{i_k},\forall i\in\mathcal{I}_k, k\in\mathcal{K}}\quad& \sum_{k=1}^{K}\sum_{i=1}^{|\mathcal{I}_k|}\text{Tr}\left(\mathbf{Q}_{i_k}\right)\tag{\ref{OP:B}}\\
\text{subject to}\quad & \tilde{R}_{i_k}\geq\psi_{i_k},\ \forall i,k\label{OP:B1}\\
&\boldsymbol{\Gamma}_{i_k}\succeq 0,\ \forall i,k \label{OP:B2}\\
&\boldsymbol{\Gamma}_{i_k}\in\mathbb{S}^{2MN\times 2MN},\ \forall i,k \label{OP:B3}\\
& \eqref{OP:A2}-\eqref{OP:A4}.
\end{align}
\end{subequations}
\begin{remark}
The two problems ~\eqref{OP:A} and~\eqref{OP:B} are equivalent and both non-convex. However, the constraint set of the optimization problem~\eqref{OP:B} is a convex set for any given $\boldsymbol{\Gamma}_{i_k},\forall i,k$.
\end{remark}
The following lemma states a desired result which simplifies the solution of~\eqref{OP:A}.
\begin{lemma}
If the optimization problem~\eqref{OP:B} is feasible for some $\boldsymbol{\Gamma}_{i_k},\forall i,k$, the solution of~\eqref{OP:B} is also achievable in the original problem~\eqref{OP:A}.
\begin{proof}
By exploiting the lower-bound in~\eqref{LBa} for the achievable rates, the non-convex $\mathcal{S}$ formed by the constraints of~\eqref{OP:A} is converted to a convex subset $\mathcal{S}^{'}$ described by the constraints of~\eqref{OP:B} which is inscribed within $\mathcal{S}$. In other words,
\begin{align}
\mathcal{S}^{'}\subset\mathcal{S}.
\end{align}
Hence any solution that is feasible in problem~\eqref{OP:B}, is feasible in~\eqref{OP:A}.
\end{proof} 
\end{lemma}
\begin{figure*}[h!]
\centering
\subfigure[\small Moderate interference channel]{
\tikzset{every picture/.style={scale=0.92}, every node/.style={scale=1}}%
\begin{tikzpicture}

\begin{axis}[%
xmin=0,
xmax=2,
xtick={  0,0.4,0.8,1.2,1.6,2},
xlabel={Rate demand (bit/cu)},
xmajorgrids,
ymin=0,
ymax=1.2,
ytick={0,0.2,0.4,0.6,0.8,1,1.2},
ylabel={Sum-power consumption},
ymajorgrids,
legend style={at={(axis cs: 0,1.2)},anchor=north west,draw=black,fill=white,legend cell align=left,opacity=0.8}
]
\addplot [color=blue,solid]
  table[row sep=crcr]{0.01	0.000844874183271367\\
  0.231111111111111	0.0211767321250653\\
  0.452222222222222	0.0480581565521711\\
  0.673333333333333	0.0841556789844329\\
  0.894444444444444	0.133665577934574\\
  1.11555555555556	0.203613451193933\\
  1.33666666666667	0.306431602442825\\
  1.55777777777778	0.465589174786966\\
  1.77888888888889	0.72757861396416\\
  2	1.18484901297139\\
  };
\addlegendentry{PGS};

\addplot [color=red,dashed,mark=asterisk,mark options={solid}]
  table[row sep=crcr]{0.01	0.00083766230419054\\
  0.231111111111111	0.0211727827154744\\
  0.452222222222222	0.0480175937824015\\
  0.673333333333333	0.0841333253306158\\
  0.894444444444444	0.133670925293553\\
  1.11555555555556	0.203646311481157\\
  1.33666666666667	0.306420438068862\\
  1.55777777777778	0.465590336493993\\
  1.77888888888889	0.72755250635681\\
  2	1.1649112443561\\
  };
\addlegendentry{IGS};

\addplot [color=green,dashed,mark=o,mark options={solid}]
  table[row sep=crcr]{0.01	0.000799313307263481\\
  0.231111111111111	0.0211747577725389\\
  0.452222222222222	0.0480411547420642\\
  0.673333333333333	0.0841327063476041\\
  0.894444444444444	0.133660759009309\\
  1.11555555555556	0.203591527002437\\
  1.33666666666667	0.306425424791834\\
  1.55777777777778	0.465387349569165\\
  1.77888888888889	0.669612201075106\\
  2	0.952683770058699\\
  };
\addlegendentry{IGS, $N=2$};

\end{axis}
\end{tikzpicture}%
\label{fig:SimoWeakInterference}
}
\subfigure[\small Strong interference channel]{
\tikzset{every picture/.style={scale=0.92}, every node/.style={scale=1}}%
\begin{tikzpicture}

\begin{axis}[%
xmin=0,
xmax=1.5,
xtick={  0,0.3,0.6,0.9,1.2,1.5},
xlabel={Rate demand (bit/cu)},
xmajorgrids,
ymin=0,
ymax=1.2,
ytick={0,0.2,0.4,0.6,0.8,1,1.2},
ylabel={Sum-power consumption},
ymajorgrids,
legend style={at={(axis cs: 0,1.2)},anchor=north west,draw=black,fill=white,legend cell align=left,opacity=0.8}
]
\addplot [color=blue,solid]
  table[row sep=crcr]{0.01	0.0231738118251299\\
  0.231111111111111	0.0398770226638273\\
  0.452222222222222	0.0731322753525026\\
  0.673333333333333	0.144106572889658\\
  0.894444444444444	0.278662338379124\\
  1.11555555555556	0.515622751518463\\
  1.33666666666667	0.989114866901235\\
  1.55777777777778	1.88928068152379\\
  1.77888888888889	3.8592694144075\\
  2	9.22598771495841\\
  };
\addlegendentry{PGS};

\addplot [color=red,dashed,mark=asterisk,mark options={solid}]
  table[row sep=crcr]{0.01	0.000806265988416181\\
  0.231111111111111	0.0235803197540695\\
  0.452222222222222	0.0593201579950363\\
  0.673333333333333	0.113411208799492\\
  0.894444444444444	0.19192484092042\\
  1.11555555555556	0.30274969263916\\
  1.33666666666667	0.456565071295557\\
  1.55777777777778	0.668134411701488\\
  1.77888888888889	0.95773565083092\\
  2	1.35240353589711\\
  };
\addlegendentry{IGS};

\addplot [color=green,dashed,mark=o,mark options={solid}]
  table[row sep=crcr]{0.01	0.000831686150923057\\
  0.231111111111111	0.0223791436594192\\
  0.452222222222222	0.0532395722695139\\
  0.673333333333333	0.0968454883043893\\
  0.894444444444444	0.157662700571261\\
  1.11555555555556	0.241754944305636\\
  1.33666666666667	0.356924329435704\\
  1.55777777777778	0.514060496755767\\
  };
\addlegendentry{IGS, $N=2$};

\end{axis}
\end{tikzpicture}%
\label{fig:SimoStrongInterference}
}
\caption{\small Two cells, two users per cell, two antennas at the base station. Minimum power required to fulfill certain rate demands which is assumed to be equal for all users. Rate demands for all users are assumed to be equal. We assume that the channels remain constant over two symbols. The symbol extension length is represented by $N$.}
\label{fig:SIMO}
\end{figure*}
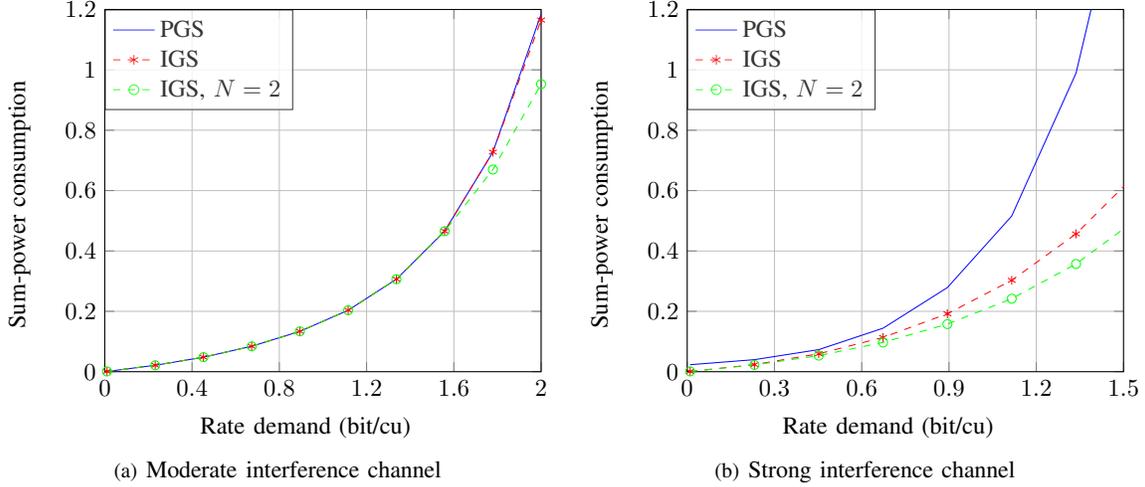
Notice that, multiple candidates exist for the set $\mathcal{S}^{'}$, due to the flexibility in choosing $\boldsymbol{\Gamma}_{i_k},\forall i,k$. Hence, a smart choice for $\boldsymbol{\Gamma}_{i_k},\forall i,k$ is necessary for the feasibility of the problem and its fast convergence. Recall that, the upper-bound gap in~\eqref{FenchelA} closes at a particular $\boldsymbol{\Gamma}_{i_k},\forall i,k$, which is hermitian positive semi-definite, i.e., $\boldsymbol{\Gamma}^{\star}_{i_k}=\mathbf{B}_{i_k}$. Hence, a realization from the positive semidefinite cone increases the possibility of non-empty feasible set. The optimization procedure is elaborated in Algorithm 1.
\begin{algorithm}\label{Alg:PowerMin}
\caption{Sum Power Minimization}
\begin{algorithmic}[1]
\State Lower-bound the second concave term in $\bar{R}_{i_k},\ \forall i,k$ as in~\eqref{FenchelA}.
\State Determine $t=1$ (iteration index)
\State Set $\epsilon$ arbitrarily low.
\State Choose symmetric positive semidefinite matrices  $\boldsymbol{\Gamma}^{(t)}_{i_k},\ \forall i,k$, which make the  problem~\eqref{OP:B} feasible
\State Define $P_{\Sigma}=\sum_{k=1}^{K}\sum_{i=1}^{|\mathcal{I}_k|}\text{Tr}\left(\mathbf{Q}_{i_k}\right)$ 

\While {$P^{(t)}_{\Sigma}-P^{(t-1)}_{\Sigma}\geq\epsilon$}
\State Solve problem~\eqref{OP:B} for the given $\boldsymbol{\Gamma}^{(t)}_{i_k},\ \forall i,k$
\State Obtain the solutions $\mathbf{Q}^{(t)}_{i_k},\ \forall i,k$
\State Calculate $P^{(t)}_{\Sigma}$
\State Obtain $\mathbf{B}^{(t)}_{i_k}$ as in~\eqref{11B}
\State Set $t=t+1$
\State Set $\boldsymbol{\Gamma}^{(t)}_{i_k}=\mathbf{B}^{(t)}_{i_k},\ \forall i,k$
\EndWhile
\State Obtain $P^{\star}_{\Sigma}=P^{(t)}_{\Sigma}$ 
\end{algorithmic}
\end{algorithm}

\begin{remark}
The rank of the sub-optimal solution captures the trade-off between multiplexing and diversity gains in the real-valued equivalent MIMO channel. For instance, a full-rank solution utilizes all real-valued equivalent MIMO degrees of freedom (DoF).  
\end{remark}
In what follows, we present the numerical results and discuss the observations and insights.
\section{Numerical Results}
We consider two channel realizations, which are representatives for moderate and strong interference regimes. By strong interference regime, we mean that the interfering channel strength is almost at the same order of the desired channels. In contrast, by moderate interference we mean that the interference channel is almost half the strength of the desired channels. For reproducibility of the results, we provide these channels in Table I for moderate and strong interference regimes when the number of antennas at the base stations is $M=2$. The first elements in the given vectors are the channel realizations for $M=1$.  Recall that $h_{kj_l}$ represents the channel from the $j$th user in the $l$th cell to the $k$th base station.\\
\begin{table*}\label{Tab:channels}
\centering
\begin{adjustbox}{max width=0.9\textwidth}
\begin{tabular}{ |c| c |c |c |c || c |c |c |c | }
\hline		
 IR& $h_{11_1}$ & $h_{12_1}$ & $h_{21_2}$ & $h_{22_2}$ &  $h_{11_2}$ & $h_{12_2}$ & $h_{21_1}$ & $h_{22_1}$  \\ \hline & & & & & & & & \\ 

   MI&$\begin{bmatrix}3.2e^{-0.72i}\\ 2.9e^{0.12i}\end{bmatrix}$ & $\begin{bmatrix}2.3e^{2.52i}\\ 3.0e^{-1.32i}\end{bmatrix}$ & $\begin{bmatrix}3.4e^{2.23i}\\ 3.1e^{0.32i}\end{bmatrix}$ & $\begin{bmatrix}3e^{-1.13i}\\ 2.9e^{0.45i}\end{bmatrix}$ &  $\begin{bmatrix}1.6e^{1.35i}\\ 1.45e^{1.23i}\end{bmatrix}$ & $\begin{bmatrix}1.15e^{0.37i}\\ 1.5e^{2.11i}\end{bmatrix}$ & $\begin{bmatrix}1.7e^{1.68i}\\ 1.55e^{0.91i}\end{bmatrix}$ & $\begin{bmatrix}1.5e^{-0.76i}\\ 1.45e^{-2.13i}\end{bmatrix}$  \\ & & & & & & & & \\ \hline & & & & & & & & \\
   SI&$-$ & $-$ & $-$ & $-$ &  $\begin{bmatrix}2.9e^{1.35i}\\ 2.7e^{1.23i}\end{bmatrix}$ & $\begin{bmatrix}2.5e^{0.37i}\\ 3.1e^{2.11i}\end{bmatrix}$ & $\begin{bmatrix}3.2e^{1.68i}\\ 2.7e^{0.91i}\end{bmatrix}$ & $\begin{bmatrix}3.1e^{-0.76i}\\ 2.4e^{-2.13i}\end{bmatrix}$  \\  & & & & & & & & \\ \hline
\end{tabular}
\end{adjustbox}
\vspace*{0.15cm}
\caption{IR: interference regime, MI: moderate interference, SI: Strong interference}
\end{table*}
The variance of the complex-valued noise at the receiver is assumed to be unity. For simulation purposes, we consider two active users in two adjacent cells. Furthermore, we consider the following cases,
\begin{enumerate}[I)]
\item only one antenna at the base station,
\item two antennas at the base station.
\end{enumerate}
The minimum sum-power consumption for achieving certain rate demands for the users is depicted in~Fig.~\ref{fig:SISO} and~Fig.~\ref{fig:SIMO} for single-antenna and two antennas base stations, respectively. One common and important observation is that, the performance of IGS outperforms PGS at low rate demands, when the interference becomes stronger. Similar observation can be made for the performance of IGS alongside symbol extensions. It is also important to notice that, by PGS higher rate demands can not be fulfilled even with very high power, however by IGS and symbol extensions high rate demands are also achievable. In the case of single-antenna base stations, having moderate interference regime, IGS improves the power efficiency of the channel, however symbol extension over two time slots is not helpful in power reduction. This can be observed in~Fig.~\ref{fig:SisoWeakInterference}. In strong interference regime, the efficiency of IGS is outstanding, moreover, by joint precoding in two time slots, the power efficiency can be even further improved.

\section{Conclusion}
In this paper, we investigated the power efficiency of IGS over an extended symbol in cellular uplink channels with inter-cell interference. The non-convex power minimization problem under rate demands turns out to be a non-convex problem (a DC program), which is efficiently solved in polynomial time. Due to the interference from the neighboring cells, the system falls into a interference-limited regime. In this case, rather than noise, the interference is the main barrier against achieving high rates. Depending on the interference regime (moderate or high interference regimes), we observed that higher rate demands are not achievable using PGS even if the mobile stations have a very high power budget. However, these rates are achievable if they utilize IGS alongside an extended symbol. Moreover, we observed that, IGS and symbol extensions are beneficial both in single antenna and multi-antenna base stations in MAC with inter-cell interference. The performance of IGS follows PGS up to particular rate demands. Hence, as future perspectives, we will analytically investigate the optimality conditions of PGS in IMAC.

\bibliographystyle{IEEEtran}
\bibliography{reference}
\end{document}